# Injection, Extraction and Matching

*M. Ferrario*
Istituto Nazionale di Fisica Nucleare – Laboratori Nazionali di Frascati, (Rome), Italy

**Abstract**
In this paper we introduce, from basic principles, the main concepts of beam focusing and transport of space charge dominated beams in high brightness accelerators using the beam envelope equation as a convenient mathematical tool. Matching conditions suitable for preserving beam quality are derived from the model for significant beam dynamics regimes. The specific case of the plasma accelerator module is also addressed.

## 1   Introduction

Light sources based on high-gain free electron lasers or future high-energy linear colliders require the production, acceleration, and transport up to the interaction point of low divergence, high charge-density electron bunches [1]. Many effects contribute in general to the degradation of the final beam quality, including chromatic effects, wake fields, emission of coherent radiation, and accelerator misalignments. Space charge effects and mismatch with focusing and accelerating devices typically contribute to emittance degradation of high charge-density beams [2], as the one required driving or being accelerated by a plasma wave; hence, proper control of beam transport, matching and extraction is the leading edge for high-quality beam acceleration in any device.

Space charge effects represent a very critical issue and a fundamental challenge for high-quality beam production and its applications. Without proper matching, significant emittance growth may occur when the beam is propagating through different stages and components owing to the large differences of transverse focusing strength. This unwanted effect is even more serious in the presence of finite energy spread leading to the so-called chromatic effect.

In this paper we introduce, from basic principles, the main concepts of beam focusing and transport in modern accelerators using the beam envelope equation as a convenient mathematical tool. Matching conditions suitable for preserving beam quality are derived from the model for significant beam dynamics regimes. A more detailed discussion of the previous topics can be found in the many classical textbooks on this subject, as listed in Refs. [3–6]. The specific case of the plasma accelerator module is also addressed.

## 2   Laminar and non-laminar beams

An ideal high-charge particle beam has orbits that flow in layers that never intersect, as occurs in a laminar fluid. Such a beam is often called a laminar beam. More precisely, an ideal laminar beam satisfies the following two conditions [6] of laminar flow:

i) All particles at a given position have identical transverse velocities. On the contrary, the orbits of two particles that start at the same position could separate and later cross each other;

ii) Assuming the beam propagates along the z axis, the magnitudes of the slopes of the trajectories in the transverse directions x and y, given by

$$x'(z) = \frac{dx}{dz} \text{ and } y'(z) = \frac{dy}{dz},$$

are linearly proportional to the displacement from the axis z of beam propagation.

Trajectories of interest in beam physics are always confined to the inside of small, near-axis regions, and the transverse momentum is much smaller than the longitudinal momentum, $p_{x,y} \ll p_z \approx p$. As a consequence, it is possible in most cases to use the small angle, or *paraxial*, approximation, which allows us to write the useful approximate expressions $x' = p_x/p_z \approx p_x/p$ and $y' = p_y/p_z \approx p_x/p$.

To help understand the features and advantages of laminar beam propagation, the following figures compare the typical behaviour of an ideal laminar and a non-laminar (or thermal) beam.

Figure 1 illustrates an example of orbit evolution of a laminar mono-energetic beam with half width $x_0$ along a simple beam line with an ideal focusing element (solenoid, magnetic quadrupoles, or electrostatic transverse fields are usually adopted to this end), represented by a thin lens located at the longitudinal coordinate $z = 0$. In an ideal lens, focusing (defocusing) forces are linearly proportional to the displacement from the symmetry axis $z$, so that the lens maintains the laminar flow of the incoming beam.

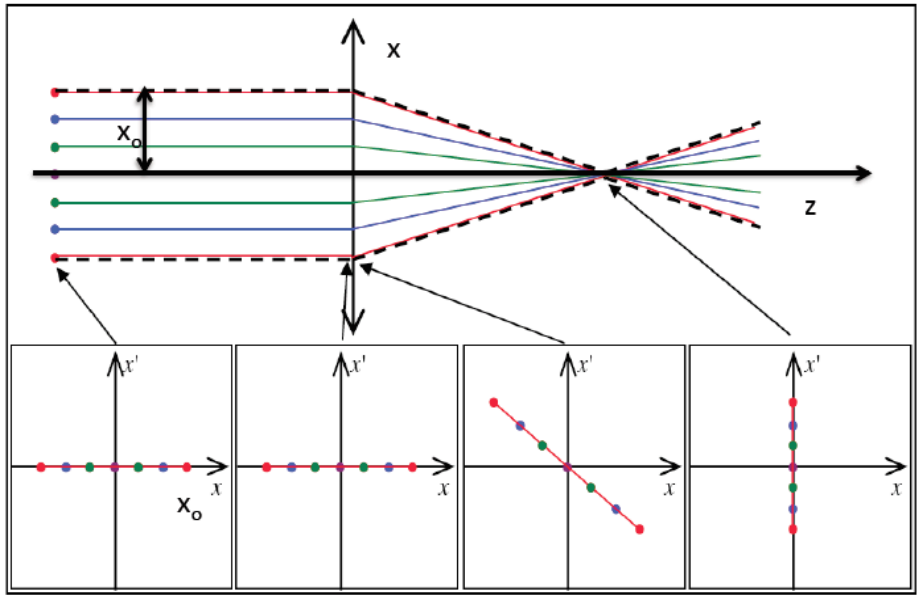

**Fig. 1:** Particle trajectories and trace space evolution of an ideal laminar beam [7]

The beam shown in Fig. 1 starts propagating completely parallel to the symmetry axis $z$, in this particular case, the particles all have zero transverse velocity. There are no orbits that cross each other in such a beam. Ignoring collisions and inner forces, such as Coulomb forces, a parallel beam could propagate an infinite distance with no change in its transverse width. When the beam crosses the ideal lens, it is transformed into a converging laminar beam. Because the transverse velocities after the linear lens are proportional to the displacement off axis, particle orbits define similar triangles that converge to a single point. After passing through the singularity at the focal point, the particles follow diverging orbits. We can always transform a diverging (or converging) beam into a parallel beam by using a lens of the proper focal length, as can be seen by reversing the propagation axis of Fig. 1.

The small boxes in the lower part of the figure depict the particle distributions in the trace space $(x, x')$, equivalent to the canonical phase space $(x, p_x \approx x'p)$ when $p$ is constant, i.e., without beam acceleration. The phase space area occupied by an ideal laminar beam is a straight segment of zero thickness. As can be easily verified, the condition that the particle distribution has zero thickness proceeds from condition 1; the segment straightness is a consequence of condition 2. The distribution of a laminar beam propagating through a transport system with ideal linear focusing elements is thus a straight segment with variable slope.

Particles in a non-laminar beam have a random distribution of transverse velocities at the same location and a spread in directions, as shown in Fig. 2. Because of the disorder of a non-laminar beam, it is impossible to focus all particles from a location in the beam toward a common point. Lenses can influence only the average motion of particles. Focal spot limitations are a major concern for a wide variety of applications, from electron microscopy to free electron lasers and linear colliders, including proper injection in a plasma accelerator. In this case the trace space plot of a non-laminar beam is no longer a straight line: the beam, as shown in the lower boxes of Fig. 2, occupies a wider area of the trace space.

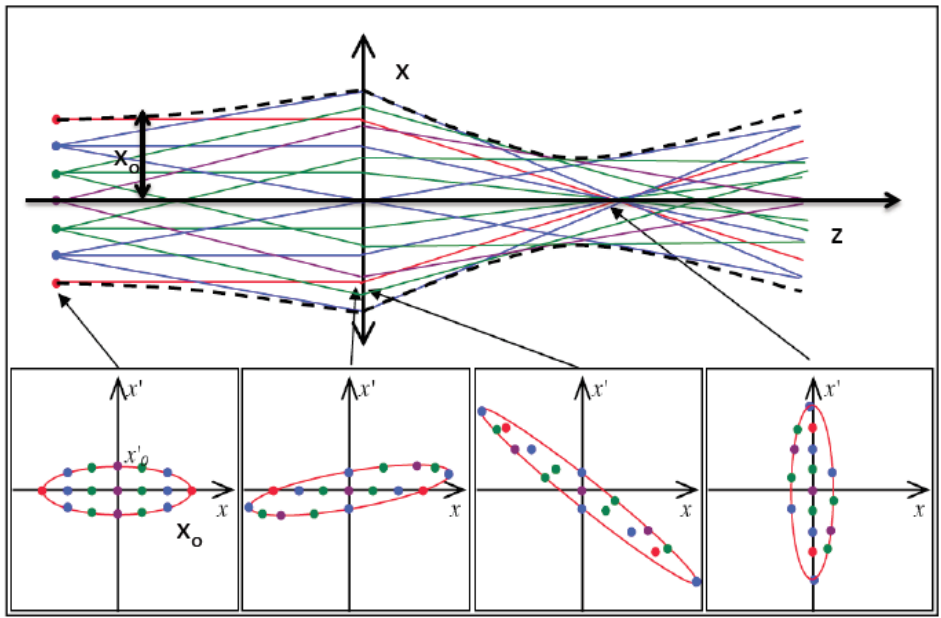

**Fig. 2:** Particle trajectories and trace space evolution of a non-laminar beam [7]

## 3 The emittance concept

From the previous discussion should appear that the trace space surface $A$ occupied by a beam is a convenient figure of merit for designating the quality of a beam. This quantity is the *geometric emittance* $\varepsilon_x$ and is usually represented by an ellipse that contains the whole particle distribution in the trace space $(x, x')$, such that $A = \pi\varepsilon_x$. An analogous definition holds for the $(y, y')$ and $(z, z')$ planes. The original choice of an elliptical shape comes from the fact that when *linear* focusing forces are applied to a beam, the trajectory of each particle in trace space lies on an ellipse, which may be called the trajectory ellipse. Being the area of the trace space, the emittance is measured in meters-radians. More often is expressed in millimetres-milliradians (mm-mrad) or, equivalently, in micrometers (μm).

The ellipse equation is written as:

$$\gamma_x x^2 + 2\alpha_x xx' + \beta_x x'^2 = \varepsilon_x \qquad (1)$$

where $x$ and $x'$ are the particle coordinates in the trace space and the coefficients $\alpha_x(z)$, $\beta_x(z)$, and $\gamma_x(z)$ are called Twiss parameters, which are related by the geometrical condition:

$$\beta_x \gamma_x - \alpha_x^2 = 1. \qquad (2)$$

As shown in Fig. 3, the beam envelope boundary $X_{max}$, its derivative $(X_{max})'$, and the maximum beam divergence $X'_{max}$, i.e., the projection on the axes $x$ and $x'$ of the ellipse edges, can be expressed as a function of the ellipse parameters:

$$\begin{cases} X_{max} = \sqrt{\beta_x \varepsilon_x} \\ (X_{max})' = -\alpha \sqrt{\dfrac{\varepsilon}{\beta}} \\ X'_{max} = \sqrt{\gamma_x \varepsilon_x} \end{cases} \qquad (3)$$

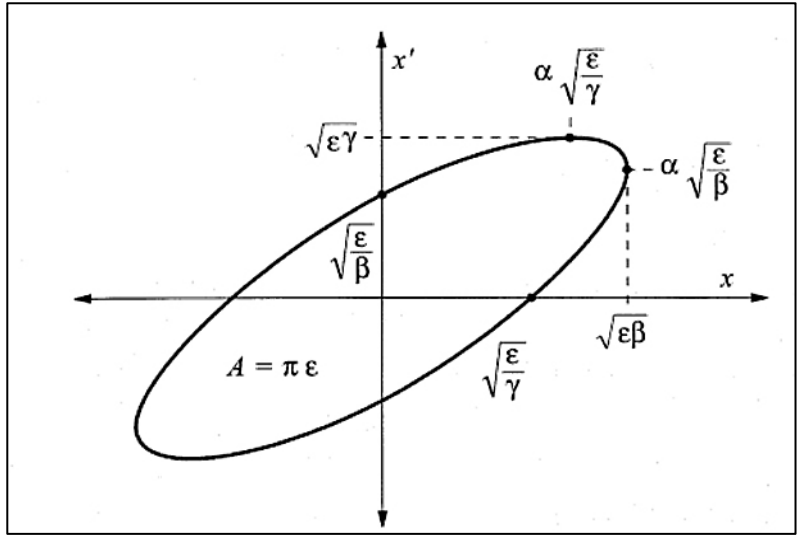

**Fig. 3:** Trace space distribution in a skewed elliptical boundary, showing the relationship of Twiss parameters to the ellipse geometry [6].

According to Liouville's theorem, the six-dimensional $(x, p_x, y, p_y, z, p_z)$ phase space volume occupied by a beam is constant, provided that there are no dissipative forces, no particles lost or created, and no Coulomb scattering among particles. As already mentioned the trace space concept is equivalent to the phase space without acceleration. Under this condition Liuoville's theorem applies also for the trace space. Moreover, if the forces in the three orthogonal directions are uncoupled, Liouville's theorem also holds for each reduced phase space surface, $(x, p_x)$, $(y, p_y)$, $(z, p_z)$, and hence the geometrical emittance remains constant in each plane [3].

Although the net trace space surface occupied by a beam is constant, non-linear field components can stretch and distort the particle distribution in the trace space, and the beam will lose its laminar behaviour. A realistic trace space distribution is often very different from a regular ellipse, as shown in Fig. 4 and a focusing system based on linear optics only may fail to accomplish a proper beam transport.

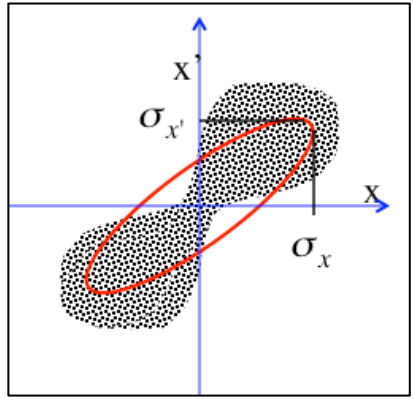

**Fig. 4:** Typical evolution of trace space distribution (black dots) under the effects of non-linear forces with the equivalent ellipse superimposed (red line).

We introduce, therefore, a definition of emittance more appropriate to measure the beam quality rather than the whole trace space area. It is often more convenient to associate a statistical definition of emittance with a generic distribution function $f(x, x', z)$ in the trace space; this is the so-called *root mean square (rms) emittance*:

$$\gamma_x x^2 + 2\alpha_x xx' + \beta_x x'^2 = \varepsilon_{x,\text{rms}}. \tag{4}$$

The rms emittance is defined such that the equivalent-ellipse projections on the $x$ and $x'$ axes are equal to the rms values of the distribution, implying the following conditions:

$$\begin{cases} \sigma_x = \sqrt{\beta_x \varepsilon_{x,\text{rms}}} \\ \sigma_{x'} = \sqrt{\gamma_x \varepsilon_{x,\text{rms}}} \end{cases}, \tag{5}$$

where:

$$\begin{cases} \sigma_x^2(z) = \langle x^2 \rangle = \int_{-\infty}^{+\infty}\int_{-\infty}^{+\infty} x^2 f(x,x',z)\,dx\,dx' \\ \sigma_{x'}^2(z) = \langle x'^2 \rangle = \int_{-\infty}^{+\infty}\int_{-\infty}^{+\infty} x'^2 f(x,x',z)\,dx\,dx' \end{cases} \tag{6}$$

are the second moments of the distribution function $f(x, x', z)$. Another important quantity that accounts for the degree of $(x, x')$ correlations is defined as:

$$\sigma_{xx'}(z) = \langle xx' \rangle = \int_{-\infty}^{+\infty} \int_{-\infty}^{+\infty} xx' f(x,x',z) \, dx \, dx' \qquad (7)$$

From Eq. (6) it also holds that:

$$\sigma'_x = \frac{\sigma_{xx'}}{\sigma_x} = -\alpha_x \sqrt{\frac{\varepsilon_{x,\text{rms}}}{\beta_x}}$$

(see also Eq. (16)), which allows us to link the correlation moment, Eq. (7), to the Twiss parameter as:

$$\sigma_{xx'} = -\alpha_x \varepsilon_{x,\text{rms}}. \qquad (8)$$

One can easily see from relations (3) and (5) that holds also: $\alpha_x = -\frac{1}{2}\frac{d\beta_x}{dz}$.

By substituting the Twiss parameter implicitly defined by Eqs. (5) and (8) into the geometrical condition (2) we obtain [5]:

$$\frac{\sigma_{x'}^2}{\varepsilon_{x,\text{rms}}} \frac{\sigma_x^2}{\varepsilon_{x,\text{rms}}} - \left(\frac{\sigma_{xx'}}{\varepsilon_{x,\text{rms}}}\right) = 1. \qquad (9)$$

Reordering the terms in Eq. (8) we get the definition of *rms emittance* in terms of the second moments of the particle trace space distribution:

$$\varepsilon_{\text{rms}} = \sqrt{\sigma_x^2 \sigma_{x'}^2 - \sigma_{xx'}^2} = \sqrt{\left(\langle x^2 \rangle \langle x'^2 \rangle - \langle xx' \rangle^2\right)}, \qquad (10)$$

where we omit, from now on, the subscript $x$ in the emittance notation: $\varepsilon_{\text{rms}} = \varepsilon_{x,\text{rms}}$. The rms emittance tells us some important information about phase space distributions under the effect of linear or non-linear forces acting on the beam. Consider, for example, an idealized particle distribution in phase space that lies on some line that passes through the origin, as illustrated in Fig. 5.

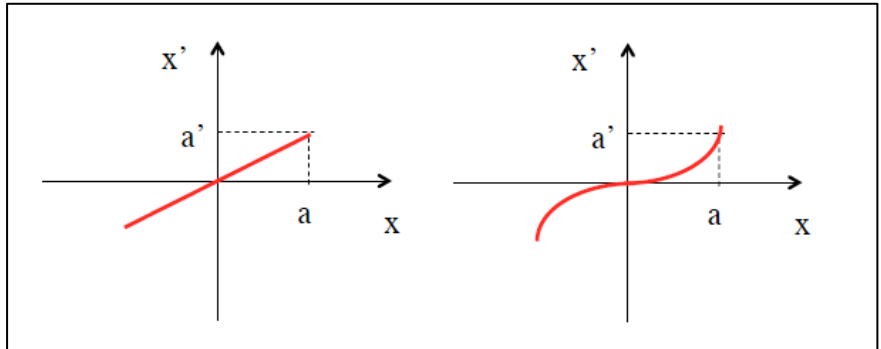

**Fig. 5:** Trace space distributions under the effect of linear (right plot) or non-linear (left plot) forces acting on the beam.

Assuming a generic $x - x'$ correlation of the type $x' = Cx^n$ and computing the rms emittance according to Eq. (10) we have

$$\varepsilon_{rms}^2 = C\sqrt{\langle x^2\rangle\langle x^{2n}\rangle - \langle x^{n+1}\rangle^2} \quad \begin{cases} n=1 \Rightarrow \varepsilon_{rms} = 0 \\ n>1 \Rightarrow \varepsilon_{rms} \neq 0 \end{cases}. \tag{11}$$

When $n = 1$, the line is straight and the rms emittance is $\varepsilon_{rms} = 0$. When $n > 1$ the relationship is non-linear, the line in phase space is curved, and the rms emittance is, in general, not zero. Both distributions have zero area. Therefore, we conclude that even when the trace space area is zero, if the distribution is lying on a curved line, its rms emittance is not zero. The rms emittance depends not only on the area occupied by the beam in trace space, but also on distortions produced by non-linear forces.

If the beam is subject to acceleration, it is more convenient to use the rms normalized emittance, for which the transverse momentum $p_x = p_z x' = m_0 c \beta \gamma x'$ is used instead of the divergence:

$$\varepsilon_{n,rms} = \frac{1}{m_o c}\sqrt{\sigma_x^2 \sigma_{p_x}^2 - \sigma_{xp_x}^2} = \frac{1}{m_o c}\sqrt{\left(\langle x^2\rangle\langle p_x^2\rangle - \langle xp_x\rangle^2\right)} = \sqrt{\left(\langle x^2\rangle\langle(\beta\gamma x')^2\rangle - \langle x\beta\gamma x'\rangle^2\right)}. \tag{12}$$

The reason for introducing a normalized emittance is that the divergences of the particles $x' = p_x/p$ are reduced during acceleration as $p$ increases. Thus, acceleration reduces the un-normalized emittance (in the trace space), but does not affect the normalized emittance that is now more properly defined by Eq. (12) in the phase space.

It is interesting to estimate the fundamental limit of the beam emittance that is set by quantum mechanics on the knowledge of the two conjugate variables $(x, p_x)$. The state of a particle is actually not exactly represented by a point, but by a small uncertainty volume of the order of $\hbar^3$ in the 6D phase space. According to the Heisenberg uncertainty relation $\sigma_x \sigma_{p_x} \geq \frac{\hbar}{2}$ one gets from Eq. (12) $\varepsilon_{n,rms}^{QM} \geq \frac{1}{2}\frac{\hbar}{m_o c} = \frac{\lambdabar_c}{2}$, where $\lambdabar_c$ is the reduced Compton wavelength. For electrons it gives: $\varepsilon_{n,rms}^{QM} \geq 1.9 \times 10^{-13} m$.

In the classical limit we see also from Eq. (12) that the single particle emittance is zero.

Assuming a small energy spread within the beam, the normalized and un-normalized emittances can be related by the approximated relation $\langle\beta\gamma\rangle\varepsilon_{rms}$. This approximation, which is often used in conventional accelerators, may be strongly misleading when adopted for describing beams with significant energy spread, like those currently produced by plasma accelerators [8]. When the correlations between the energy and transverse positions are negligible (as in a drift without collective effects), Eq. (12) can be written as:

$$\varepsilon_{n,rms}^2 = \langle\beta^2\gamma^2\rangle\langle x^2\rangle\langle x'^2\rangle - \langle\beta\gamma\rangle^2\langle xx'\rangle^2. \tag{13}$$

Considering now the definition of relative energy spread $\sigma_\gamma^2 = \frac{\langle\beta^2\gamma^2\rangle - \langle\beta\gamma\rangle^2}{\langle\beta\gamma\rangle^2}$ which can be inserted into Eq. (13) to give:

$$\varepsilon_{n,\mathrm{rms}}^2 = \left\langle \beta^2\gamma^2 \right\rangle \sigma_\gamma^2 \left\langle x^2 \right\rangle \left\langle x'^2 \right\rangle + \left\langle \beta\gamma \right\rangle^2 \left( \left\langle x^2 \right\rangle \left\langle x'^2 \right\rangle - \left\langle xx' \right\rangle^2 \right). \tag{14}$$

Assuming relativistic particles ($\beta = 1$), we get:

$$\varepsilon_{n,\mathrm{rms}}^2 = \left\langle \gamma^2 \right\rangle \left( \sigma_\gamma^2 \sigma_x^2 \sigma_{x'}^2 + \varepsilon_{\mathrm{rms}}^2 \right). \tag{15}$$

If the first term in the parentheses is negligible, we find the conventional approximation of the normalized emittance as $\left\langle \gamma \right\rangle \varepsilon_{\mathrm{rms}}$. For a conventional accelerator, this might generally be the case. Considering, for example, beam parameters for the SPARC_LAB photoinjector [9]: at 5 MeV the ratio between the first and the second term is $\sim 10^{-3}$ while at 150 MeV it is $\sim 10^{-5}$. Conversely, using typical beam parameters at the plasma–vacuum interface, the first term is of the same order of magnitude as for conventional accelerators at low energies; however, owing to the rapid increase of the bunch size outside the plasma ($\sigma_{x'} \sim$ mrad) and the large energy spread ($\sigma_\gamma > 1\%$), it becomes predominant compared with the second term after a drift of a few millimetres. *Therefore, the use of approximated formulas when measuring the normalized emittance of plasma accelerated particle beams is inappropriate* [10].

## 4    The root mean square envelope equation

We are now interested in following the evolution of the particle distribution during beam transport and acceleration. One can use the collective variable defined in Eq. (6), the second moment of the distribution termed the rms beam envelope, to derive a differential equation suitable for describing the rms beam envelope dynamics [11]. To this end, let us compute the first and second derivative of $\sigma_x$ [4]:

$$\begin{aligned}
\frac{d\sigma_x}{dz} &= \frac{d}{dz}\sqrt{\left\langle x^2 \right\rangle} = \frac{1}{2\sigma_x}\frac{d}{dz}\left\langle x^2 \right\rangle = \frac{1}{2\sigma_x}2\left\langle xx' \right\rangle = \frac{\sigma_{xx'}}{\sigma_x} \\
\frac{d^2\sigma_x}{dz^2} &= \frac{d}{dz}\frac{\sigma_{xx'}}{\sigma_x} = \frac{1}{\sigma_x}\frac{d\sigma_{xx'}}{dz} - \frac{\sigma_{xx'}^2}{\sigma_x^3} = \frac{1}{\sigma_x}\left(\left\langle x'^2 \right\rangle + \left\langle xx'' \right\rangle\right) - \frac{\sigma_{xx'}^2}{\sigma_x^3} = \frac{\sigma_{x'}^2 + \left\langle xx'' \right\rangle}{\sigma_x} - \frac{\sigma_{xx'}^2}{\sigma_x^3}
\end{aligned} \tag{16}$$

Rearranging the second derivative in Eq. (16), we obtain a second-order non-linear differential equation for the beam envelope evolution:

$$\sigma_x'' = \frac{\sigma_x^2 \sigma_{x'}^2 - \sigma_{xx'}^2}{\sigma_x^3} + \frac{\left\langle xx'' \right\rangle}{\sigma_x} \tag{17}$$

or, in a more convenient form, using the rms emittance definition of Eq. (10):

$$\sigma_x'' - \frac{1}{\sigma_x}\left\langle xx'' \right\rangle = \frac{\varepsilon_{\mathrm{rms}}^2}{\sigma_x^3}. \tag{18}$$

In Equation (18), the emittance term can be interpreted physically as an outward pressure on the beam envelope produced by the rms spread in trajectory angle, which is parameterized by the rms emittance. Let us now consider, for example, the simple case with $\langle xx''\rangle = 0$, describing a beam drifting in free space. The envelope equation reduces to:

$$\sigma_x^3 \sigma_x'' = \varepsilon_{rms}^2 \qquad (19)$$

With initial conditions $\sigma_0$, $\sigma'_0$ at $z_0$, depending on the upstream transport channel, Eq. (19) has a hyperbolic solution:

$$\sigma(z) = \sqrt{\left(\sigma_o + \sigma'_o (z - z_o)\right)^2 + \frac{\varepsilon_{rms}^2}{\sigma_o^2}(z - z_o)^2} \;. \qquad (20)$$

Considering the case of a beam at waist ($\langle xx'\rangle = 0$) with $\sigma'_0 = 0$, using Eq. (5), the solution Eq. (20) is often written in terms of the $\beta$ function as:

$$\sigma(z) = \sigma_0 \sqrt{1 + \left(\frac{z - z_0}{\beta_w}\right)^2} \;. \qquad (21)$$

This relation indicates that without any external focusing element the beam envelope increases from the beam waist by a factor $\sqrt{2}$ with a characteristic length $\beta_w = \sigma_0^2 / \varepsilon_{rms}$, as shown in Fig. 6.

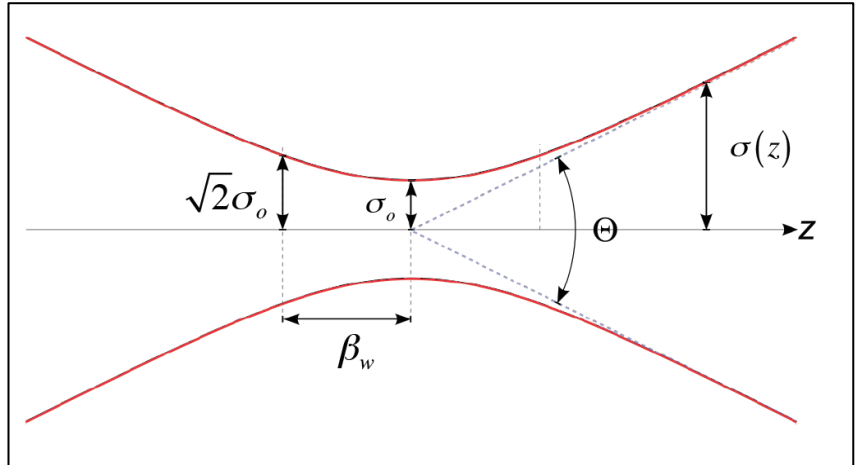

**Fig. 6:** Schematic representation of the beam envelope behaviour near the beam waist

At waist holds also the relation $\varepsilon_{rms}^2 = \sigma_{o,x}^2 \sigma_{o,x'}^2$ that can be inserted in Eq. (20) to get $\sigma_x^2(z) = \sigma_{o,x'}^2 (z - z_o)^2$. Under this condition, Eq. (15) can be written as:

$$\varepsilon_{n,rms}^2(z) = \langle \gamma^2 \rangle \left( \sigma_\gamma^2 \sigma_{x'}^4 (z - z_0)^2 + \varepsilon_{rms}^2 \right)$$

showing that beams with large energy spread and divergence undergo a significant normalized emittance growth even in a drift of length ($z - z_0$) [8, 12].

Notice also that the solution Eq. (21) is exactly analogous to that of a Gaussian light beam for which the beam width $w = 2\sigma_{ph}$ increases away from its minimum value at the waist $w_0$ with characteristic length $Z_R = \pi w_0^2 / \lambda$ (Rayleigh length) [4]. This analogy suggests that we can identify an equivalent emittance of a photon beam as $\varepsilon_{ph} = \lambda/4\pi$.

For an effective transport of a beam with finite emittance is mandatory to make use of some external force providing beam confinement in the transport or accelerating line. The term $\langle xx'' \rangle$ accounts for external forces when we know $x''$ given by the single particle equation of motion:

$$\frac{dp_x}{dt} = F_x . \tag{22}$$

Under the paraxial approximation $p_x \ll p = \beta\gamma mc$, the transverse momentum $p_x$ can be written as $p_x = px' = \beta\gamma m_0 c x'$, so that:

$$\frac{dp_x}{dt} = \frac{d}{dt}(px') = \beta c \frac{d}{dz}(px') = F_x \tag{23}$$

and the transverse acceleration results to be:

$$x'' = -\frac{p'}{p} x' + \frac{F_x}{\beta c p} . \tag{24}$$

It follows that:

$$\langle xx'' \rangle = -\frac{p'}{p}\langle xx' \rangle + \frac{\langle xF_x \rangle}{\beta c p} = \frac{p'}{p}\sigma_{xx'} + \frac{\langle xF_x \rangle}{\beta c p} . \tag{25}$$

Inserting Eq. (25) into Eq. (18) and recalling $\sigma_x' = \sigma_{xx'}/\sigma_x$, the complete rms envelope equation is:

$$\sigma_x'' + \frac{p'}{p}\sigma_x' - \frac{1}{\sigma_x}\frac{\langle xF_x \rangle}{\beta c p} = \frac{\varepsilon_{n,rms}^2}{\gamma^2 \sigma_x^3} , \tag{26}$$

where we have included the approximate normalized emittance $\varepsilon_{n,rms} = \gamma\varepsilon_{rms}$. Notice that the effect of longitudinal accelerations appears in the rms envelope equation as an oscillation damping term, called 'adiabatic damping', proportional to $p'/p$. The term $\langle xF_x \rangle$ represents the moment of any external transverse force acting on the beam, such as that produced by a focusing magnetic channel.

# 5     External forces

Let's now consider the case of external linear force acting on the beam in the form $F_x = \mp kx$. It can be focusing or defocusing, according to the sign. The moment of the force is:

$$\langle xF_x \rangle = \mp k \langle x^2 \rangle = \mp k \sigma_x^2 \qquad (27)$$

and the envelope equation becomes:

$$\sigma_x'' + \frac{\gamma'}{\gamma}\sigma_x' \mp k_{ext}^2 \sigma_x = \frac{\varepsilon_{n,rms}^2}{\gamma^2 \sigma_x^3}, \qquad (28)$$

where we have explicitly used the momentum definition $p = \gamma mc$ for a relativistic particle with $\beta \approx 1$ and defined the wavenumber:

$$k_{ext}^2 = \frac{k}{\gamma m_0 c^2}$$

Typical focusing elements are quadrupoles and solenoids [3] and more recently Active Plasma Lenses have shown interesting capabilities [20]. The magnetic quadrupole field is given in Cartesian coordinates by

$$\begin{cases} B_x = B_0 \dfrac{y}{d} = B_0' y \\ B_y = B_0 \dfrac{x}{d} = B_0' x \end{cases}, \qquad (29)$$

where $d$ is the pole distance and $B_0'$ is the field gradient. The force acting on the beam is $\vec{F}_\perp = qv_z B_0' (y\hat{j} - x\hat{i})$ and, when $B_0$ is positive, is focusing in the $x$ direction and defocusing in the $y$ direction. The focusing strength is:

$$k_{quad} = \frac{qB_0'}{\gamma m_0 c} = k_{ext}^2$$

In a solenoid the focusing strength is given by:

$$k_{sol} = \left(\frac{qB_0}{2\gamma m_0 c}\right)^2 = k_{ext}^2$$

Notice that the solenoid is always focusing in both directions, an important property when the cylindrical symmetry of the beam must be preserved. However, being a second-order quantity in $\gamma$, it is more effective at low energy.

In an Active Plasma Lens (APL) the focusing force is provided by the azimuthal magnetic field $B_\theta$ generated by a discharge current $I_c$ in a short (cm scale) discharge capillary [20 - 23], with typical radius $R_c$ less than 1 mm. The resulting magnetic field is equivalent to the one generated by a current flowing in a wire and has in principle perfect cylindrical symmetry thus providing focusing in both planes at the same time. The focusing strength is given by

$$k_{APL} = \frac{eB'_\vartheta}{\gamma mc} = \frac{2I_c}{\gamma I_A R_c^2} = k_{ext}^2 \quad ,$$

where $I_A = 4\pi\varepsilon_0 m_0 c^3/e = 17$ kA is the Alfvén current for electrons. For example with a discharge current of 450 A in a capillary of 300 μm radius the expected field reaches the remarkable gradient of 1 kT/m. APL concept in currently under investigation in many laboratories to optimize its performances and remove possible non-linearity in the focusing field. Its simplicity, compactness and effectiveness make this concept very attractive especially in the framework of Plasma Accelerator development.

It is interesting to consider the case of a uniform focusing channel without acceleration described by the rms envelope equation

$$\sigma''_x + k_{ext}^2 \sigma_x = \frac{\varepsilon_{rms}^2}{\sigma_x^3} \quad . \tag{30}$$

By substituting $\sigma_x = \sqrt{\beta_x \varepsilon_{rms}}$ into Eq. (30) one obtains an equation for the 'betatron function' $\beta_x(z)$ that is independent of the emittance term:

$$\beta''_x + 2k_{ext}^2 \beta_x = \frac{2}{\beta_x} + \frac{\beta'^2_x}{2\beta_x} \quad . \tag{31}$$

Equation (31) contains just the transport channel focusing strength and, being independent of the beam parameters, suggests that the meaning of the betatron function is to account for the transport line characteristic. The betatron function reflects external forces from focusing magnets and it is highly dependent on the particular arrangement of the quadrupole magnets. The equilibrium, or matched, solution of Eq. (31) is given by $\beta_{eq} = \frac{1}{k_{ext}} = \frac{\lambda_\beta}{2\pi}$, as can be easily verified. This result shows that the matched $\beta_x$ function is simply the inverse of the focusing wavenumber or, equivalently, is proportional to the 'betatron wavelength' $\lambda_\beta$. The corresponding envelope equilibrium condition, i.e., a stationary solution of Eq. (30), is given by:

$$\sigma_{eq,x} = \sqrt{\frac{\varepsilon_{rms}}{k_{ex}}} \quad .$$

It is interesting to note that in analogy with the kinetic theory of gases we can define the beam temperature in a transverse direction at equilibrium and without correlations as:

$$k_B T_{beam,x} = \gamma m_0 \langle v_x^2 \rangle = \frac{\sigma_{p_x}^2}{\gamma m_0} = m_0 c^2 \frac{\varepsilon_{n,rms}^2}{\gamma \sigma_{eq,x}^2} = \gamma m_0 \beta^2 c^2 \frac{\varepsilon_{rms}}{\beta_{eq,x}} \quad ,$$

where $k_B$ is the Boltzmann constant and we have used Eq. (12), showing that the conditions for a "cold" beam are typically: low emittance, low energy, high betatron function.

By means of the beam temperature concept one can also define the beam emittance at the source called the thermal emittance. Assuming that electrons are in equilibrium with the cathode temperature $T_c = T_{beam}$ and $\gamma=1$, the thermal emittance is given by $\varepsilon_{th,rms}^{cat} = \sigma_x \sqrt{\frac{k_B T_c}{m_0 c^2}}$ which, per unit rms beam spot size at the cathode, is $\varepsilon_{th,rms} = 0.3$ μm/mm at $T_c = 2500$ K. For comparison, in a photocathode illuminated by a laser pulse with photon energy $\hbar\omega$ the expression for the variance of the transverse momentum of the emitted electrons is given by $\sigma_{p_x} = \sqrt{\frac{m_0}{3}(\hbar\omega - \phi_{\text{eff}})}$, where $\phi_{\text{eff}} = \phi_w - \phi_{\text{Schottky}}$, $\phi_w$ being the material work function and $\phi_{\text{Schottky}}$ the Schottky work function [19]. The corresponding thermal emittance is $\varepsilon_{\text{th,rms}}^{\text{ph}} = \sigma_x \sqrt{\frac{\hbar\omega - \phi_{\text{eff}}}{3m_0 c^2}}$ that, with the typical parameters of a Copper photocathode illuminated by a UV laser, gives a thermal emittance per unit spot size of about 0.5 μm/mm.

# 6   Space charge forces

Another important force acting on the beam is the one produced by the beam itself due to the internal Coulomb forces. The net effect of the Coulomb interaction in a multiparticle system can be classified into two regimes [3]:

i) *collisional regime*, dominated by binary collisions caused by close particle encounters;

ii) *collective regime* or *space charge regime*, dominated by the self-field produced by the particles' distribution, which varies appreciably only over large distances compared with the average separation of the particles.

A measure for the relative importance of collisional versus collective effects in a beam with particle density *n* is the relativistic *Debye length*:

$$\lambda_D = \sqrt{\frac{\varepsilon_0 \gamma^2 k_B T_b}{e^2 n}} \ . \tag{32}$$

As long as the Debye length remains small compared with the particle bunch transverse size, the beam is in the space charge dominated regime and is not sensitive to binary collisions. Smooth functions for the charge and field distributions can be used in this case, and the space charge force can be treated as an external applied force. The space charge field can be separated into linear and non-linear terms as a function of displacement from the beam axis. The linear space charge term defocuses the beam and leads to an increase in beam size. The non-linear space charge terms also increase the rms emittance by distorting the phase space distribution. Under the paraxial approximation of particle motion, we can consider the linear component alone. We shall see next that the linear component of the space charge field can also induce emittance growth when longitudinal correlations along the bunch are taken into account.

For a bunched beam of uniform charge distribution in a cylinder of radius *R* and length *L*, carrying a current $\hat{I}$ and moving with longitudinal velocity $v_z = \beta c$, the linear component of the longitudinal and transverse space charge field are given approximately by [13]:

$$E_z(\zeta) = \frac{\hat{I}L}{2\pi\varepsilon_0 R^2 \beta c} h(\zeta) \tag{33}$$

$$E_r(r,\zeta) = \frac{\hat{I}r}{2\pi\varepsilon_0 R^2 \beta c} g(\zeta) \ . \tag{34}$$

The field form factors are described by the functions:

$$h(\zeta) = \sqrt{A^2 + (1-\zeta)^2} - \sqrt{A^2 + \zeta^2} - |1-\zeta| + |\zeta| \tag{35}$$

$$g(\zeta) = \frac{(1-\zeta)}{2\sqrt{A^2 + (1-\zeta)^2}} + \frac{\zeta}{2\sqrt{A^2 + \zeta^2}}, \tag{36}$$

where $\zeta = z/L$ is the normalized longitudinal coordinate along the bunch, $\zeta = 0$ being the bunch tail, and $A = R/\gamma L$ is the beam aspect ratio. The field form factors account for the variation of the fields along the bunch and outside the bunch for $\zeta<0$ and $\zeta>L$. As $\gamma$ increases, $A \to 0$, hence $g(\zeta) \to 1$ and $h(\zeta) \to 0$, thus showing that space charge fields mainly affect transverse beam dynamics. The definition of A shows also that an energy increase corresponds to a bunch lengthening in the moving frame $L' = \gamma L$, leading to a vanishing longitudinal field component, as in the case of a continuous beam in the laboratory frame.

To evaluate the force acting on the beam one must account also for the azimuthal magnetic field associated with the beam current, that in cylindrical symmetry is given by $B_\vartheta = \frac{\beta}{c} E_r$. Thus, the Lorentz force acting on each single particle is given by:

$$F_r = e(E_r - \beta c B_\vartheta) = e(1 - \beta^2) E_r = \frac{eE_r}{\gamma^2} . \tag{37}$$

The attractive magnetic force, which becomes significant at high velocities, tends to compensate for the repulsive electric force. Therefore, space charge defocusing is primarily a non-relativistic effect and decreases as $\gamma^{-2}$.

To include space charge forces in the envelope equation, let us start writing the space charge forces produced by the previous fields in Cartesian coordinates:

$$F_x = \frac{e\hat{I}x}{2\pi\gamma^2 \varepsilon_0 \sigma_x^2 \beta c} g(\zeta) . \tag{38}$$

Then, to compute the moment of the force, we need to know:

$$x'' = \frac{F_x}{\beta c p} = \frac{eIx}{8\pi\varepsilon_0 \gamma^3 m_0 \beta^3 c^3 \sigma_x^2} = \frac{k_{sc}(\zeta)}{(\beta\gamma)^3 \sigma_x^2}, \tag{39}$$

where we have introduced the generalized beam perveance:

$$k_{sc}(\zeta) = \frac{\hat{I}}{2I_A} g(\zeta), \tag{40}$$

where $I_A$ is the Alfvén current for electrons. Notice that in this case the perveance in Eq. (40) explicitly depends on the position $\zeta$ along the bunch. We can now calculate the term that enters the envelope equation for a relativistic beam:

$$\langle xx'' \rangle = \frac{k_{sc}}{\gamma^3 \sigma_x^2} \langle x^2 \rangle = \frac{k_{sc}}{\gamma^3} \tag{41}$$

leading to the complete envelope equation:

$$\sigma_x'' + \frac{\gamma'}{\gamma}\sigma_x' + k_{ext}^2 \sigma_x = \frac{\varepsilon_{n,rms}^2}{\gamma^2 \sigma_x^3} + \frac{k_{sc}}{\gamma^3 \sigma_x} \quad . \tag{42}$$

From the envelope equation Eq. (42), we can identify two regimes of beam propagation: *space charge dominated* and *emittance dominated*. A beam is space charge dominated as long as the space charge collective forces are largely dominant over the emittance pressure. In this regime, the space charge force produces a laminar propagation of the beam, as one can verify by integrating Eq. (39): the magnitudes of the slope of the trajectories in the transverse direction x (and y) is linearly proportional to the initial displacement $x_0$ from the axis z of beam propagation.

A measure of the relative importance of space charge effects versus emittance pressure is given by the *laminarity parameter*, defined as the ratio between the space charge term and the emittance term:

$$\rho = \frac{\hat{I}}{2 I_A \gamma} \frac{\sigma^2}{\varepsilon_n^2} \quad . \tag{43}$$

When $\rho$ greatly exceeds unity, the beam behaves as a laminar flow and transport and acceleration require a careful tuning of focusing and accelerating elements to keep laminarity. Correlated emittance growth is typical in this regime, which can be made reversible if proper beam matching conditions are fulfilled, as discussed in the following. When $\rho < 1$, the beam is emittance dominated (thermal regime) and space charge effects can be neglected. The transition to the thermal regime occurs when $\rho \approx 1$, corresponding to the transition energy:

$$\gamma_{tr} = \frac{\hat{I}}{2 I_A} \frac{\sigma^2}{\varepsilon_n^2} \quad . \tag{44}$$

For example, a beam with $\hat{I}$ = 100 A, $\varepsilon_n$ = 1 μm, and $\sigma$ = 300 μm is leaving the space charge dominated regime and is entering the thermal regime at the transition energy of 131 MeV. From this example, one may conclude that the space charge dominated regime is typical of low-energy beams. Actually, for such applications as linac-driven free electron lasers, peak currents exceeding kA are required. Space charge effects may recur if bunch compressors are active at higher energies and a new energy threshold with higher $\hat{I}$ must be considered.

# 7   Correlated emittance oscillations

When longitudinal correlations within the bunch are important, like the one induced by space charge, beam envelope evolution is generally dependent also on the coordinate along the bunch $\zeta$. In this case, the bunch should be considered as an ensemble of *n* longitudinal slices of envelope $\sigma_s(z,\zeta)$, whose evolution can be computed from *n* slice envelope equations equivalent to Eq. (42), provided that the bunch parameters refer to each single slice: $\gamma_s$, $\gamma'_s$, $k_{sc,s} = k_{sc}g(\zeta)$. Correlations within the bunch may cause emittance oscillations that can be evaluated, once an analytical or numerical solution [13] of the slice envelope equation is known, by using the following correlated emittance definition:

$$\varepsilon_{rms,cor} = \sqrt{\langle \sigma_s^2 \rangle \langle \sigma_s'^2 \rangle - \langle \sigma_s \sigma_s' \rangle^2} \quad , \tag{45}$$

where the average is performed over the entire slice ensemble, assuming uniform charge distribution within each slice. In the simplest case of a two-slice model, the previous definition reduces to:

$$\varepsilon_{\text{rms,cor}} = |\sigma_1 \sigma_2' - \sigma_2 \sigma_1'| , \qquad (46)$$

which represents a simple and useful formula for an estimation of the emittance scaling [14].

The total normalized rms emittance is given by the superposition of the correlated and uncorrelated terms [15-18] as:

$$\varepsilon_{\text{rms,cor}} = \langle \gamma \rangle \sqrt{\varepsilon_{\text{rms}}^2 + \varepsilon_{\text{rms,cor}}^2} . \qquad (47)$$

An interesting example to consider here, showing the consequences of non-perfect beam matching, is the propagation of a beam in the space charge dominated regime nearly matched to an external focusing channel, as illustrated in Fig. 7.

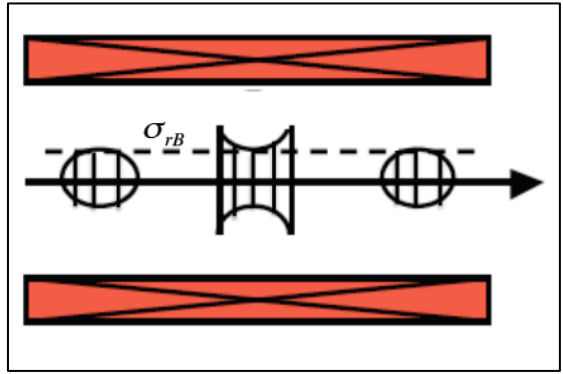

**Fig. 7:** Schematic representation of a nearly matched beam in a long solenoid. The dashed line represents the reference slice envelope matched to the Brillouin flow condition. The other slice envelopes are oscillating around the equilibrium solution.

To simplify our computations, we can neglect acceleration, as in the case of a simple beam transport line made by a long solenoid ($k^2_{\text{ext}} = k_{\text{sol}}$). The envelope equation for each slice, indicated as $\sigma_s$, reduces to:

$$\sigma_s'' + k_{\text{ext}}^2 \sigma_s = \frac{k_{\text{sc,s}}}{\gamma^3 \sigma_s} . \qquad (48)$$

A stationary solution corresponding to slice propagation with constant envelope, called *Brillouin flow*, is given by:

$$\sigma_{r,B} = \frac{1}{k_{\text{ext}}} \sqrt{\frac{\hat{I} g(\zeta)}{2\gamma^3 I_A}} , \qquad (49)$$

where the local dependence of the current $\hat{I}_s = \hat{I} g(\zeta)$ within the bunch has been explicitly indicated. This solution represents the matching conditions for which the external focusing completely balances the internal space charge force. Unfortunately, since $k_{\text{ext}}$ has a slice-independent constant value, the Brillouin matching condition is different for each slice and usually cannot be achieved at the same time for all of the bunch slices. Assuming that there is a reference slice perfectly matched (49) with an envelope $\sigma_{r,B}$ and negligible beam energy spread, the matching condition for the other slices can be written as:

$$\sigma_{sB} = \sigma_{rB} + \frac{\sigma_{rB}}{2}\left(\frac{\delta I_s}{\hat{I}}\right) \qquad (50)$$

with respect to the reference slice. Considering a slice with a small perturbation $\delta_s$ with respect to its own equilibrium Eq. (50) in the form:

$$\sigma_s = \sigma_{s,B} + \delta_s \qquad (51)$$

and substituting into Eq. (48), we can obtain a linearized equation for the slice offset:

$$\delta_s'' + 2k_{ext}^2 \delta_s = 0, \qquad (52)$$

which has a solution given by:

$$\delta_s = \delta_0 \cos\left(\sqrt{2}k_{ext}z\right), \qquad (53)$$

where $\delta_0 = \sigma_{so} - \sigma_{sB}$ is the amplitude of the initial slice mismatch, which we assume, for convenience, is the same for all slices. Inserting Eq. (53) into Eq. (51) we get the perturbed solution:

$$\sigma_s = \sigma_{s,B} + \delta_0 \cos\left(\sqrt{2}k_{ext}z\right). \qquad (54)$$

Equation (54) shows that slice envelopes oscillate together around the equilibrium solution with the same frequency for all slices ($\sqrt{2}k_{ext}$, often called the plasma frequency) dependent only on the external focusing forces. This solution represents a collective behaviour of the bunch, similar to that of the electrons subject to the restoring force of ions in plasma. Using the two-slice model and Eq. (54), the emittance evolution Eq. (46) results in:

$$\varepsilon_{rms,cor} = \frac{1}{4}k_{ext}\sigma_{rB}\left|\frac{\Delta I}{\hat{I}}\delta_0 \sin(k_{ext}z)\right|, \qquad (55)$$

where $\Delta I = \hat{I}_1 - \hat{I}_2$. Notice that, in this simple case, envelope oscillations of the mismatched slices induce correlated emittance oscillations that periodically return to zero, showing the reversible nature of the correlated emittance growth. It is, in fact, the coupling between transverse and longitudinal motion induced by the space charge fields that allows reversibility. With proper tuning of the transport line length or of the focusing field, one can compensate for the transverse emittance growth.

At first, it may seem surprising that a beam with a single charge species can exhibit plasma oscillations, which are characteristic of plasmas composed of two-charge species. However, the effect of the external focusing force can play the role of the other charge species, providing the necessary restoring force that is the cause of such collective oscillations, as shown in Fig. 8. The beam can actually be considered as a single-component, relativistic, cold plasma.

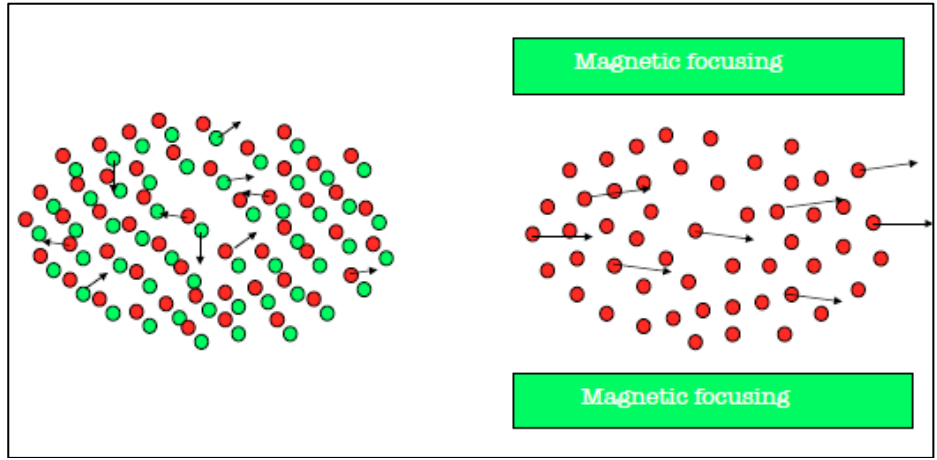

**Fig. 8:** The restoring force produced by the ions (green dots) in plasma may cause electron (red dots) oscillations around the equilibrium distribution. In a similar way, the restoring force produced by a magnetic field may cause beam envelope oscillations around the matched envelope equilibrium.

It is important to bear in mind that beams in linacs are also different from plasmas in some important respects [5]. One is that beam transit time through a linac is too short for the beam to reach thermal equilibrium. Also, unlike plasma, the Debye length of the beam may be larger than, or comparable to, the beam radius, so shielding effects may be incomplete.

## 8    Matching conditions in a Plasma Accelerator

The concepts developed for the beam transport in the previous sections can be now applied to the case of a plasma accelerator [24] giving important information about the critical topic of beam-plasma matching conditions. To this end we introduce a simplified model for the plasma and for the resulting fields acting on the beam in order to be able to write an envelope equation for the accelerated beam. In this section we are interested in the case of external injection of particles in a plasma wave that could be excited by a short intense laser pulse [24,25] or by a driving electron beam [26,27].

A very simplified model for the plasma channel behind the driving pulse is illustrated in Fig. 9.

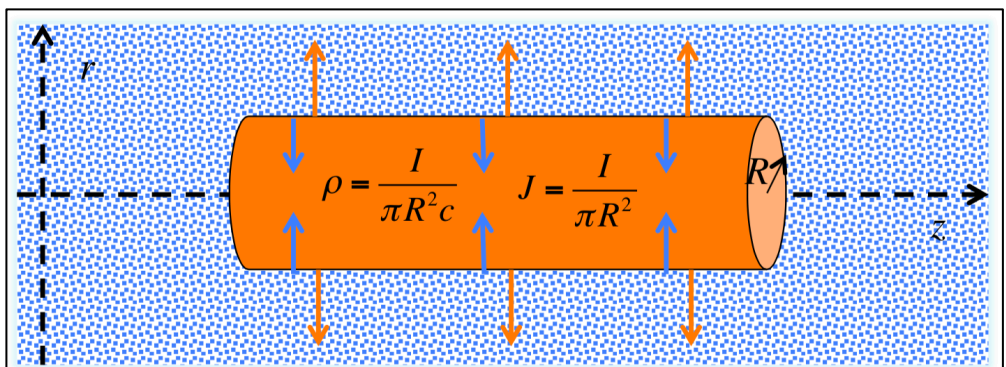

**Fig. 9:** Schematic representation of a relativistic cylindrical bunch of radius R, charge density ρ and current density J, propagating in an ionized gas background of opposite charge (blue dots). Orange (blue) arrows represent the defocusing (focusing) space charge force produced by the electron beam (ion background).

In this model a relativistic witness electron bunch is propagating in a ionized gas background of opposite charge, a simplified configuration similar to the one produced in a capillary discharge [23] where a plasma oscillation (charge separation) has beam excited by a driving pulse. The transverse electric and magnetic fields experienced by a particle can be evaluated by considering a cylindrical beam space charge model [3] that gives:

$$\begin{cases} E_r = \dfrac{I(1-f_e)}{2\pi\varepsilon_o R^2 v} r & \text{for } r \leq R \\ E_r = \dfrac{I(1-f_e)}{2\pi\varepsilon_o v} \dfrac{1}{r} & \text{for } r > R \end{cases} \qquad \begin{cases} B_\vartheta = \mu_o \dfrac{I(1-f_m)}{2\pi R^2} r & \text{for } r \leq R \\ B_\vartheta = \mu_o \dfrac{I(1-f_m)}{2\pi} \dfrac{1}{r} & \text{for } r > R \end{cases} \quad . \tag{56}$$

The factor $f_e$ accounts for a stationary charge distribution of opposite sign that results in a partial neutralization of the space charge of the primary beam particles. The $f_m$ factor accounts for a partial magnetic neutralization produced by the plasma return current. The short electron pulse propagating through the plasma indeed produces a time-varying azimuthal magnetic field $B_\theta$ that generate a longitudinal electric field $E_z$. This $E_z$ field generates an ion current in the opposite direction to the incoming electron beam and the magnetic field associated with this return current is opposite to the magnetic field of the primary electron beam. Consequently one gets a partial magnetic neutralization of which the $f_m$ factor accounts.

Using the fields (56) the Lorentz force acting on the electrons can be written as:

$$F_r = e(E_r - \beta c B_\vartheta) = \frac{eE_r}{\gamma^2}\left(1 - \gamma^2 f_e + \beta^2 \gamma^2 f_m\right) \quad . \tag{57}$$

Following the same procedure leading to equation (42) we can now model this case with a more general envelope equation for a relativistic beam:

$$\sigma'' + \frac{\gamma'}{\gamma}\sigma' + \frac{k_{ext}^2}{\gamma}\sigma = \frac{2I\left(1 - \gamma^2 f_e + \gamma^2 f_m\right)}{I_A \gamma^3 \sigma} + \frac{\varepsilon_n^2}{\gamma^2 \sigma^3} \tag{58}$$

that reduces to (42) when $f_e = f_m = 0$ as must be in the case of in vacuum propagation. It is worth noticing that the new "space charge" term on the right hand side of this equation can contribute to focus or defocus the beam depending on the sign of the term between brackets:

$$\left(1 - \gamma^2 f_e + \beta^2 \gamma^2 f_m\right) \begin{cases} < 0 & \Rightarrow \text{focusing effect} \\ = 0 & \Rightarrow \text{screening effect} \\ > 0 & \Rightarrow \text{defocusing effect} \end{cases} \tag{59}$$

Let consider the charge screening effect of the plasma background with particle density $n_p$, by defining $f_e = n_p/n_e$, where $n_e$ is the bunch particle density and $f_m = 0$. We also do not include any external focusing element so that $k_{ext} = 0$. The envelope equation (58) can be recast as:

$$\sigma'' + \frac{\gamma'}{\gamma}\sigma' + \frac{2In_p}{I_A \gamma n_e \sigma} = \frac{2I}{I_A \gamma^3 \sigma} + \frac{\varepsilon_n^2}{\gamma^2 \sigma^3} \tag{60}$$

or by recalling the definition of the Alfven current $I_A$ and of the beam current $I=ecn_e\pi\sigma^2$ it can be written in a more familiar form as:

$$\sigma'' + \frac{\gamma'}{\gamma}\sigma' + \frac{k_p^2}{2\gamma}\sigma = \frac{2I}{I_A\gamma^3\sigma} + \frac{\varepsilon_n^2}{\gamma^2\sigma^3} \ , \tag{61}$$

where $k_p^2 = \frac{e^2 n_p}{\varepsilon_o mc^2}$ is the usual plasma wave number that accounts for the background ion focusing effect.

An equilibrium solution of the previous equation has not yet been found and numerical integration is the only way to solve it. Nevertheless some simplification is still possible and an approximated matching condition it exists. As one can see there are two focusing terms, the adiabatic damping and the ion focusing, and two defocusing terms, the emittance pressure and the space charge effects. With the typical beam parameters of a plasma accelerator: 1 kA peak current, 1 µm normalized emittance, injection energy $\gamma_o$=300 and spot size about 5 µm, the laminarity parameter (43) results to be $\rho < 1$, i.e., the beam is emittance dominated. We have also to neglect the adiabatic damping term, setting $\gamma'$=0, i.e., no acceleration. This approximation is quite strong but it will allow us to find at least a proper matching condition that holds at the entrance and at the exit of the plasma column. It follows that the envelope equation can be approximated by the reduced expression:

$$\sigma'' + \frac{k_p^2}{2\gamma}\sigma = \frac{\varepsilon_n^2}{\gamma^2\sigma^3} \ . \tag{62}$$

Looking for an equilibrium solution of (62) we obtain:

$$\sigma_{eq} = \sqrt[4]{\frac{2}{\gamma}}\sqrt{\frac{\varepsilon_n}{k_p}} \ , \tag{63}$$

where $\gamma$ is the normalized beam energy at the entrance (exit) of the plasma. Before injection in the plasma accelerator, the beam has to be focused to the matched spot size given by (63) to prevent envelope oscillations that may cause emittance growth and an enhancement of betatron radiation emission. It has been proposed [16, 28-31] to shape the plasma density profile in order to gently capture the beam by means of the increasing ion focusing effect. For example, by slowly varying the density $n_P(z)$ with a plasma ramp at the entrance of the plasma column. Equation (63) still holds with $k_P(z)$ dependent on the ramp profile and thus the beam envelope can be gently matched to the accelerating plasma channel.

## Acknowledgements

I wish to thank A. Cianchi, E. Chiadroni, R. Pompili, J.B. Rosenzweig, A.R. Rossi, A. Bacci, and L. Serafini for the many helpful discussions and suggestions.